\RequirePackage{silence}
\documentclass[preprint]{aastex701}
\usepackage{amsmath}
\usepackage{amssymb}
\usepackage{tipa}
\usepackage{CJKutf8}
\usepackage{combelow}
\usepackage{graphicx}
\usepackage{subcaption}

\begin{document}

\title{Optical Colors as Compressed Spectral Information for Trans-Neptunian Objects}

\author[0000-0001-7737-6784]{Hsing~Wen~Lin ({\normalfont\begin{CJK*}{UTF8}{bkai}
林省文\end{CJK*}})}
\affiliation{Department of Physics, University of Michigan, Ann Arbor, MI 48109, USA}
\affiliation{Michigan Institute for Data and AI in Society, University of Michigan, Ann Arbor, MI 48109, USA}
\affiliation{Department of Physics, Case Western Reserve University, Cleveland, OH 44106, USA}
\email{hsing-wen.lin@case.edu}

\author[0000-0002-2486-1118]{Larissa Markwardt}
\affiliation{University of Auckland, Science Centre 303, 38 Princes St, Auckland Central, Auckland, 1010, New Zealand}
\email{larissa.markwardt@auckland.ac.nz}

\author[0000-0003-4827-5049]{Kevin J. Napier} 
\affiliation{Center for Astrophysics $\mid$ Harvard \& Smithsonian, Cambridge, MA 02138, USA}
\affiliation{Michigan Institute for Data and AI in Society, University of Michigan, Ann Arbor, MI 48109, USA}
\affiliation{Department of Physics, University of Michigan, Ann Arbor, MI 48109, USA}
\email{kjnapier@umich.edu}

\author[0000-0002-8167-1767]{Fred C. Adams}
\affiliation{Department of Physics, University of Michigan, Ann Arbor, MI 48109, USA}
\affiliation{Department of Astronomy, University of Michigan, Ann Arbor, MI 48109, USA}
\email{fca@umich.edu}

\author[0000-0002-1226-3305]{Renu Malhotra}
\affiliation{Lunar and Planetary Laboratory, The University of Arizona, Tucson, AZ 85721, USA}
\email{malhotra@arizona.edu}

\author[0000-0001-6942-2736]{David W. Gerdes}
\affiliation{Department of Physics, Case Western Reserve University, Cleveland, OH 44106, USA}
\affiliation{Department of Physics, University of Michigan, Ann Arbor, MI 48109, USA}
\affiliation{Department of Astronomy, University of Michigan, Ann Arbor, MI 48109, USA}
\email{dwg50@case.edu}

\begin{abstract}
Broadband optical colors have long been recognized as tracers of the surface spectral properties of trans-Neptunian objects (TNOs), yet their relationship to near-infrared (NIR) spectra has not been quantitatively characterized. We construct an empirical optical-NIR spectral manifold from JWST spectroscopy of 40 TNOs and 8 Neptune Trojans and interpret broadband photometry as compressed representations of this manifold. Using the spectral manifold as a prior, the model conditions on observed broadband optical colors to infer a posterior distribution in the spectral latent space. We then evaluate filter combinations through posterior entropy and Kullback-Leibler information gain and found that the $gr$ colors provide broad localization on the manifold, while additional filters substantially reduce posterior uncertainty. Independently defined optical color populations map coherently onto the major compositional classes of the spectral manifold, providing a physical interpretation of previously identified color groups. The learned manifold also enables the inverse mapping back to optical color space. Sampling and decoding the manifold reproduces the observed topology of optical color space, indicating that these color populations arise naturally as low-dimensional projections of a continuous optical-NIR compositional manifold rather than isolated empirical clusters. The forward and inverse mappings provide a unified interpretation of optical colors as compressed representations of an optical-NIR spectral manifold. As JWST expands the manifold and Rubin maps millions of objects onto it, this framework provides an innovated approach to population-scale compositional studies and the discovery of rare surface types.
\end{abstract}


\section{Introduction} 
\label{sec:intro}
Trans-Neptunian objects (TNOs) exhibit a remarkably broad range of optical colors \citep{Luu1996AJ, Tegler1998, Peixinho2012, Peixinho2015, Pike2017, Schwamb2019, Fraser2023, Bernardinelli2025, Ferreira2025MNRAS}. These colors are usually interpreted as signatures of varied surface compositions, as they are low-resolution measurements of an object's reflectance spectrum. However, optical spectra of TNOs are often relatively featureless and are dominated by broad continuum slopes \citep{Barucci2008, Alvarez-Candal2008}, making it difficult to identify specific surface materials from optical data alone. More diagnostic compositional information is generally found at near-infrared (NIR) wavelengths, where absorption features associated with ices and complex organics become accessible \citep{Barucci2008, Merlin2012}. The detailed physical meaning of TNO color diversity has therefore remained difficult to establish: optical photometry is abundant, whereas NIR  spectroscopy has historically been sparse and often limited to the brightest objects.

JWST has begun to change this situation by providing a large and relatively homogeneous NIR spectroscopic view of TNO surfaces \citep{Souza-Feliciano2024, disco2024, DePra2024, discocentaur2024}. an be classified into three surface compositional classes, namely, water-rich, CO$_2$-rich, and organic-rich surfaces \citep{Holler2025}; the organic-rich class further divided into methanol-rich and methanol-poor spectral subtypes \citep{Brunetto2025}. The availability of these JWST spectral samples allows us to ask whether and how optical colors encode NIR spectral information. Recent studies by \citet{Markwardt2025} and \citet{Bernardinelli2025} have begun to establish relations between optical colors and TNO NIR spectral morphology, indicating that broadband color carries information about spectral structure. If optical color populations map coherently onto NIR spectral classes, then some fraction of the information contained in NIR spectra must already be encoded, though incompletely, in optical photometry. 

\citet{Ferreras2023} have examined how to interpret astronomical spectra as information carriers and to quantify their information content from an information-theoretic perspective, by treating spectra as probability distributions and measuring their Shannon entropy. Data-driven generative forward models that map physical parameters or lower-dimensional observables onto spectra have a long history in stellar spectroscopy, from The Cannon \citep{cannon} to The Payne \citep{payne}, enabling both label inference and spectral reconstruction from a shared latent representation.
\citet{Doorenbos2024ApJ} have demonstrated that spectroscopic information can be inferred from lower-dimensional observations such as broadband imaging or photometry in the context of galaxies. Motivated by these developments, in this work we treat optical colors as incomplete measurements of an underlying optical--NIR spectral manifold and infer posterior distributions in the spectral latent space rather than assigning objects directly to fixed spectral labels. This framework allows us to quantify the information content of different optical filter sets, map optical color populations onto NIR spectral classes, and identify candidates for rare or transitional surface types.

This paper is organized as follows. Section~\ref{sec:method} describes the spectral sample, optical-color extension, probabilistic reconstruction framework, and information metrics used in the analysis. Section~\ref{sec:results} presents the information content of optical filter sets, the mapping of optical color groups onto the spectral manifold, and generating the optical color distribution from the spectral manifold. Section~\ref{sec:dis} discusses the interpretation and potential applications of these results, and Section~\ref{sec:sum} is summary of this work.

\section{The Spectral Manifold in Latent Space} \label{sec:method}

\subsection{Spectral Sample and Optical-Color Extension} \label{sec:sample}
Our empirical spectral sample is constructed from two JWST NIR spectroscopic datasets with available optical photometry. The first consists of 40 TNOs from the DiSCo-TNO survey (\citealp{disco2024}, \dataset[DOI: 10.17909/r2zp-r280]{https://doi.org/10.17909/r2zp-r280}), and the second consists of eight Neptune Trojans (\citealp{Markwardt2025}, \dataset[DOI: 10.17909/j66s-pv96]{https://doi.org/10.17909/j66s-pv96}). Together these objects span the major spectral classes currently identified among typical trans-Neptunian populations and provide paired optical colors and NIR reflectance spectra for constructing the empirical spectral manifold. We also restrict the training sample to objects with both JWST spectroscopy and corresponding optical photometry. This does not represent every published outer Solar System spectrum. Objects lacking homogeneous optical photometry, as well as populations with substantially different evolutionary histories (e.g., Centaurs, \citealp{discocentaur2024}) or distinct surface processes (e.g., dwarf planets, \citealp{Emery2024}), were not included. The resulting sample is therefore designed to be representative of the dominant compositional diversity of typical TNOs while providing a consistent training set for the color-to-spectrum inference problem.

\begin{figure}
\centering
\includegraphics[width=1\columnwidth]{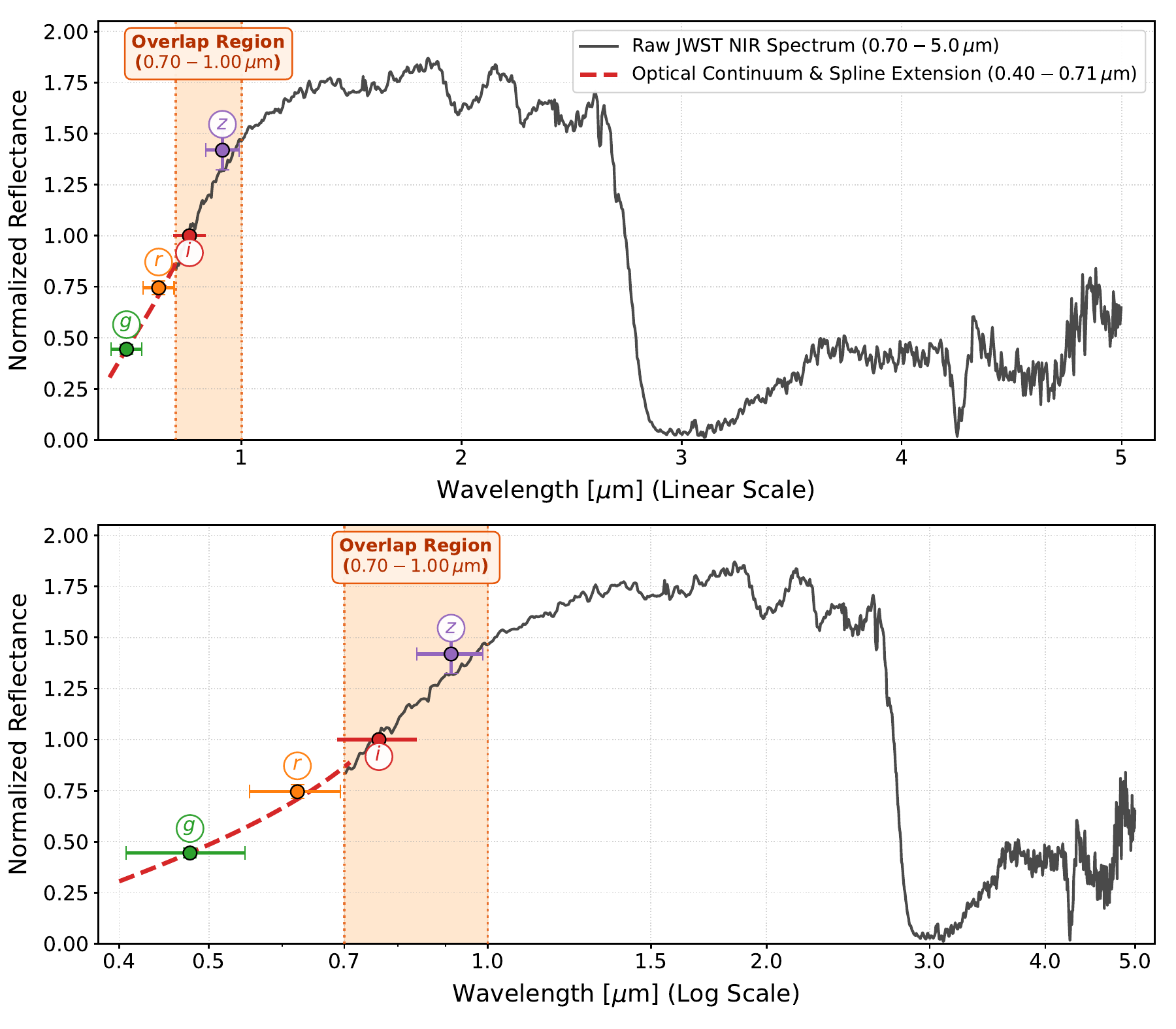}
\caption{Example construction of a continuous optical--NIR reflectance spectrum from joining optical colors with the JWST NIR spectrum. The top and bottom panels present the construction in linear and logarithmic wavelength scales, respectively. The dark solid line is the raw JWST NIR spectrum from 0.70 to 5.1 $\mu$m, the circular markers with error bars represent the $griz$ photometric constraints and the red dashed curve shows the optical continuum using spline extension.}
\label{fig:opt+NIR}
\end{figure}

The published JWST spectra cover approximately $0.75$--$5.1~\mu{\rm m}$. To construct a continuous optical--NIR reflectance spectrum, we extend each spectrum to optical wavelengths using the available visible colors\footnote{For the DiSCo-TNO sample, we use the spectral slopes, $S'$, reported by \citet{disco2024}. For the Neptune Trojans, we use optical colors from \citet{2018AJ....155...56J}, \citet{2013AJ....145...96P}, \citet{Bolin2023}, and \citet{Markwardt2023}.}. The optical colors are converted into relative reflectance constraints, which define the visible continuum between approximately $0.4$ and $0.75~\mu{\rm m}$ (g to i bands). We then join this optical continuum smoothly to the short-wavelength end of the JWST spectrum using the wavelength region where the optical extension and NIR spectrum overlap. This transition is implemented with a cubic spline, producing a continuous reflectance spectrum from $0.4$ to $5.1~\mu{\rm m}$. Figure~\ref{fig:opt+NIR} illustrates this procedure for an object, showing how the optical color constraints are connected to the JWST spectrum to produce the final combined optical--NIR spectrum.

This construction treats optical photometry as a low-resolution constraint on the same reflectance spectrum, rather than as a separate set of classification features. Because the optical region ($0.4$--$1.0~\mu{\rm m}$) spans a much smaller fraction of the full wavelength range, we resample the joined spectra on a logarithmic wavelength grid (Figure~\ref{fig:opt+NIR}, lower panel) before constructing the latent representation. The logarithmic grid increases the relative weight of the short-wavelength optical region while preserving the broad NIR absorption and continuum structures that dominate the spectral diversity.

\subsection{Probabilistic Inference on the Spectral Manifold}
\label{sec:reconstruction}

After preprocessing, each object is represented as a continuous optical--NIR reflectance spectrum over $0.4$--$5.1~\mu{\rm m}$. We then construct a probabilistic optical--NIR spectral manifold following \citet{Lin2026}. Principal component analysis (PCA) provides a low-dimensional latent representation of the spectra, while kernel density estimation (KDE) defines an empirical prior over physically plausible regions of this latent space. The PCA basis and KDE prior jointly define the optical--NIR spectral manifold and its empirical density.

The main difference from \citet{Lin2026} is that the conditioning data are now broadband optical colors rather than NIR photometry. Given an observed optical color vector $\mathbf{c}$, the model infers a posterior probability distribution over the latent spectral manifold,
\begin{equation}
p(\mathbf{z}\,|\,\mathbf{c}),
\end{equation}
where $\mathbf{z}$ denotes the PCA latent coordinates of the optical--NIR spectrum. Rather than assigning each object to a single point in latent space, the Bayesian framework propagates observational uncertainty into a posterior distribution over physically plausible spectra. This posterior representation forms the basis for all subsequent analyses in this work, including information-theoretic metrics (section~\ref{sec:metrics}), probabilistic compositional inference (section~\ref{sec:info}), and the bidirectional mapping between optical colors and the spectral manifold (section~\ref{sec:C2S} and ~\ref{sec:S2C}).

Samples drawn from $p(\mathbf{z}\,|\,\mathbf{c})$ can be decoded through the PCA basis to generate corresponding optical--NIR spectra. Throughout this work, spectral decoding is used primarily for visualization and for generating synthetic optical color distributions from the learned manifold, rather than as the primary scientific product.

Following \citet{Lin2026}, we use the first ten principal components to represent and decode the spectral manifold. Information-theoretic quantities and divergence measures are computed using the first five principal components, which capture the dominant compositional structure while reducing sensitivity to higher-order interpolation noise and small-scale spectral variations.

\subsection{Information Metrics in Latent Space}
\label{sec:metrics}

The posterior distribution $p(\mathbf{z}\,|\,\mathbf{c})$ provides more information than a single reconstructed spectrum or a discrete spectral label. It describes not only where an object is likely to lie in the optical--NIR spectral manifold, but also how uncertain that location is, and how close it is to known spectral families or rare spectral-types. We therefore characterize each posterior distribution using two complementary information metrics in latent space.

\begin{figure}
\centering
\includegraphics[width=1\columnwidth]{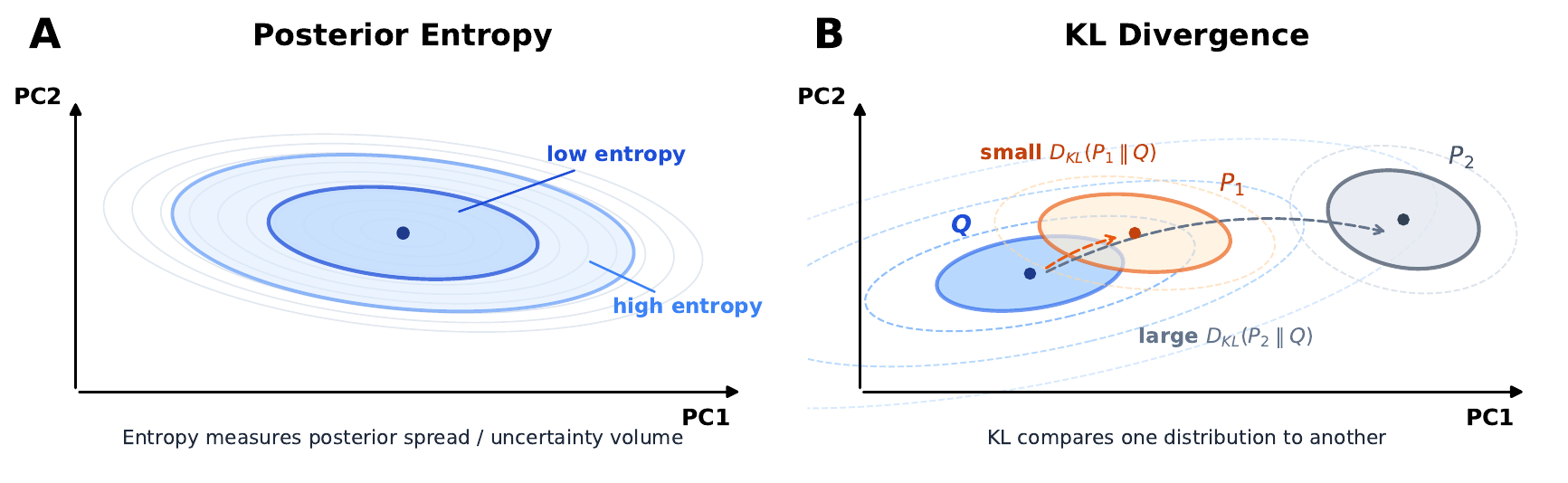}
\caption{
Geometric interpretation of the latent-space information metrics used in this work. In each panel, the background contours represent the empirical spectral prior in latent space. Entropy measures the effective volume of the posterior distribution, and KL divergence compares two latent distributions.%; surprisal measures rarity under the empirical spectral prior; and Mahalanobis distance measures covariance-scaled proximity to a reference object, centroid, or spectral phenotype. These metrics are complementary: uncertainty, distribution mismatch, rarity, and similarity are distinct properties of posterior spectral inference.
}
\label{fig:metric_schematic}
\end{figure}

First, we use the Shannon entropy of the posterior distribution (hereafter, entropy, see Figure~\ref{fig:metric_schematic} left panel) as a measure of reconstruction uncertainty. For a posterior distribution $p(\mathbf{z}\,|\,\mathbf{c})$, the entropy is

\begin{equation}
H[p(\mathbf{z}\,|\,\mathbf{c})]
= -\int p(\mathbf{z}\,|\,\mathbf{c})
\log p(\mathbf{z}\,|\,\mathbf{c})\,d\mathbf{z}.
\label{eq:entropy}
\end{equation}
In this context, entropy quantifies the effective volume of latent spectral space allowed by the observed optical colors. A low-entropy posterior indicates that the colors strongly constrain the possible spectra, whereas a high-entropy posterior indicates that multiple spectral morphologies remain compatible with the same optical measurements

Second, we use the Kullback-Leibler divergence (hereafter, KL divergence, see Figure~\ref{fig:metric_schematic} right panel) to compare latent distributions. For two distributions $p(\mathbf{z})$ and $q(\mathbf{z})$, the KL divergence is

\begin{equation}
D_{\rm KL}(p\Vert q)
= \int p(\mathbf{z})
\log
\frac{p(\mathbf{z})}{q(\mathbf{z})}
\,d\mathbf{z}.
\label{eq:kl}
\end{equation}
This quantity measures the additional information required of describing samples drawn from $p$ using the reference distribution $q$. In this work, KL divergence is used to compare posterior distributions with reference distributions in the spectral latent. For individual objects, the posterior distribution inferred from the observed optical colors is compared with the empirical latent distribution of each spectroscopic compositional class, providing a probabilistic basis for spectral classification. For population-level analyses, KL divergence is similarly used to compare the latent distributions of optically defined color groups with those of empirical near-infrared spectral families. In practice, all distributions are estimated from latent samples using kernel density estimation. KL divergence therefore serves as a distribution-to-distribution similarity metric rather than a point-to-point distance or a calibrated classification probability.

%Third, we use surprisal to quantify how unusual a given object or posterior sample is under the empirical spectral prior. If $p_{\rm prior}(\mathbf{z})$ is the KDE estimate of the latent prior density, the surprisal is

%\begin{equation}
%S(\mathbf{z})
%= -\log p_{\rm prior}(\mathbf{z}).
%\label{eq:surprisal}
%\end{equation}
%Objects with high surprisal occupy low-density regions of the empirical spectral prior and are therefore natural candidates for rare or under-sampled spectral-types. This metric is especially useful for identifying objects in low-density tails of otherwise broad spectral families.

%Finally, we use Mahalanobis distance to quantify proximity to reference objects, class centroids, or rare spectral phenotypes in a covariance-scaled latent space. For a latent point $\mathbf{z}$, a reference point or centroid $\boldsymbol{\mu}$, and covariance matrix $\Sigma$, the Mahalanobis distance is the Euclidean distance after scaling by the covariance structure:

%\begin{equation}
%d_{\rm M}(\mathbf{z},\boldsymbol{\mu})=
%\sqrt{
%(\mathbf{z}-\boldsymbol{\mu})^{T}
%5\Sigma^{-1}
%(\mathbf{z}-\boldsymbol{\mu})
%}.
%\label{eq:mahalanobis}
%\end{equation}
%Unlike surprisal, which measures rarity with respect to the full empirical prior, Mahalanobis distance measures similarity to a specified reference. It is therefore useful for identifying candidates that are not global outliers but lie close to known transition-region objects.

Figure~\ref{fig:metric_schematic} provides a geometric interpretation of these quantities in latent space, illustrating how entropy characterizes the spread, or uncertainty of the posterior distribution, whereas KL divergence measures its similarly to the empirical spectral classes. The two information metrics allow us to move beyond direct label assignment and instead analyze optical colors as probabilistic constraints on the geometry of the spectral manifold.

\section{Results: From Colors to Spectra and Back Again} \label{sec:results}

\subsection{Information Content of Optical Filter Sets}
\label{sec:info}
We first quantify how much information different optical filter sets provide about the joint optical-NIR spectral manifold using leave-one-out cross-validation on the 48-object spectral sample described above. For each held-out object, we perform the optical-to-latent reconstruction without using that object's spectrum in the training set and infer the posterior distribution using only the selected optical colors.

The posterior entropy measures how uncertain the inferred latent position remains after conditioning on a given filter set, and the information gain is defined as the KL divergence between this posterior and the empirical spectral prior,
\begin{equation}
{\rm IG}(\mathbf{c}) =
D_{\rm KL}\left[p(\mathbf{z}|\mathbf{c})\,||\,p_{\rm prior}(\mathbf{z})\right].
\end{equation}
This quantity measures how much the optical observations move the inferred latent distribution away from the prior expectation. Posterior entropy and information gain quantify complementary aspects of the reconstruction: how localized the posterior is, and how strongly it differs from the prior.

\begin{figure}
\centering
\includegraphics[width=.8\columnwidth]{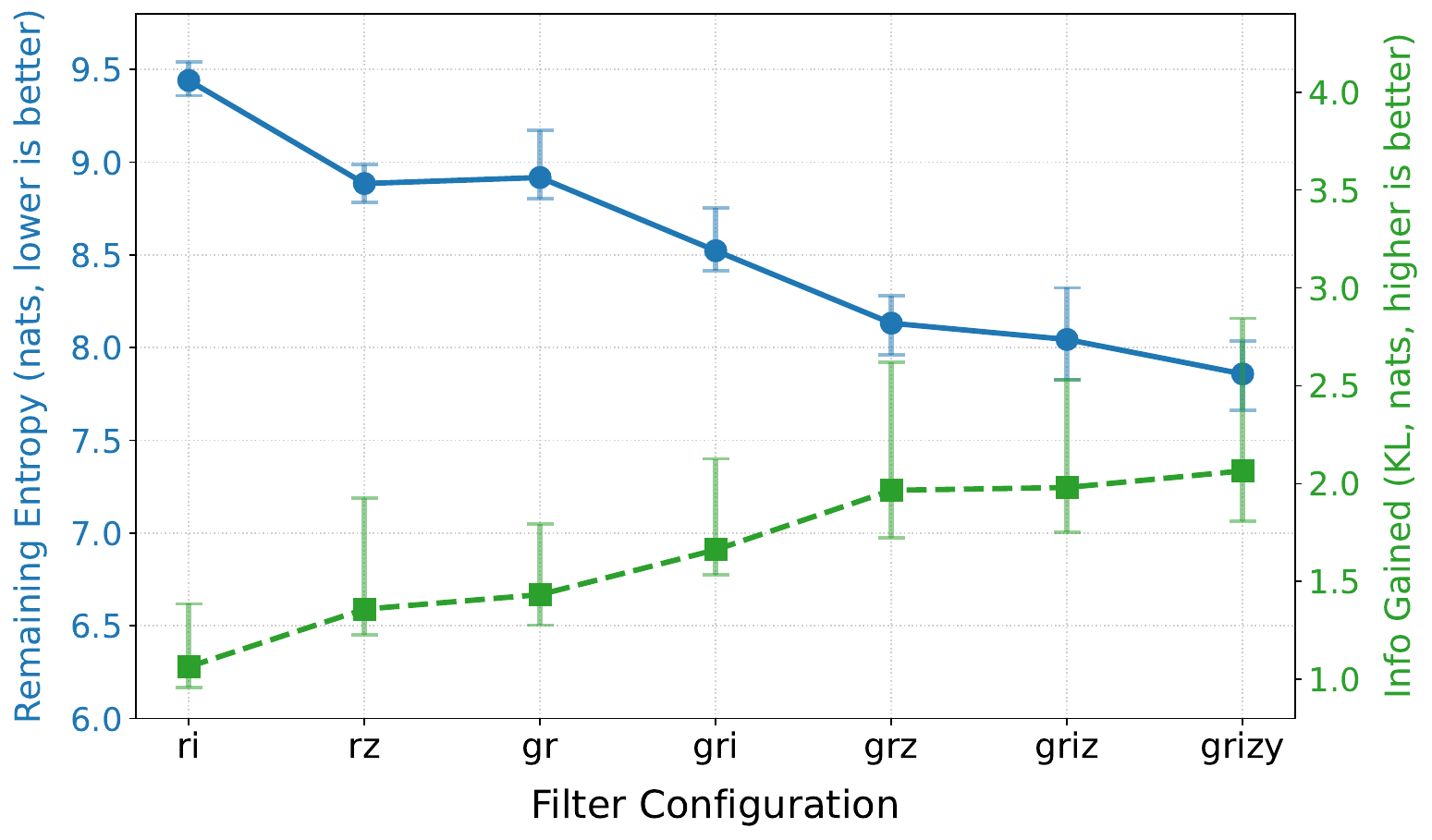}
\caption{
Expected information content of different Rubin/LSST-like optical filter configurations. The blue solid curve shows the expected remaining posterior entropy (lower is better), and the green dashed curve shows the expected information gain (higher is better). Points indicate the median across all 48 benchmark spectra under 100 Monte Carlo 5\% photometric uncertainty realizations per object. Vertical error bars denote the 16th-84th percentile interval across the target population.}
\label{fig:filter_information}
\end{figure}

Figure~\ref{fig:filter_information} compares the expected posterior entropy and information gain for seven Rubin/LSST filter configurations: $ri$, $rz$, $gr$, $gri$, $grz$, $griz$, and $grizy$. For each leave-one-out realization, the posterior distribution of the excluded object is estimated using 100 Monte Carlo samples, and the reported values represent averages over all leave-one-out trials.

The results demonstrate that the information content of photometry is strongly wavelength-dependent. Among the two-band configurations, $gr$ and $rz$ provide comparable and substantial constraints on the optical--NIR spectral manifold, with information gains of approximately $1.4$ nats.
The importance of both the $g$ and $z$ bands becomes particularly evident in three-filter combinations. Adding $g$ to $rz$ increases the information gain by $\sim0.61$ nats and reduces the posterior entropy by $\sim0.75$ nats. Similarly, adding $z$ to $gr$ increases the information gain by $\sim0.53$ nats and reduces the posterior entropy by $\sim0.79$ nats.

Among the three-band configurations, $grz$ provides lower posterior entropy and higher information gain than $gri$, demonstrating that constraints on the underlying spectral manifold depend not only on the number of bands but also on their wavelength coverage.
This comparison does not imply that the $i$ band is unimportant: $gri$ still provides a substantial improvement over $gr$, while the relative utility of $i$ and $z$ can also depend on the signal-to-noise and observing efficiency of a given survey. 

Adding further redder bands produces progressively smaller gains. The $griz$ configuration performs similarly to $grz$, indicating that the incremental information provided by $i$ becomes modest once the $g-r-z$ baseline is established. The full $grizy$ configuration yields the lowest expected posterior entropy, but its information gain is similar to that of $griz$, suggesting that the $y$ band primarily provides a modest refinement of the posterior rather than a substantial shift in its inferred spectral constraints.

Overall, the results show a clear pattern of diminishing returns: either $gr$ or $rz$ provides a strong initial constraint on an object's location on the spectral manifold, and adding a third band substantially tightens this constraint, whereas further bands mainly provide incremental refinement. In particular, the substantial improvements from $rz$ to $grz$ and from $gr$ to $grz$ highlight the information contributed by the $g$ and $z$ bands, respectively, in constraining an object's location on the optical--NIR spectral manifold. This wavelength dependence is an important consideration when optimizing multi-band survey strategies.

As a practical application of this information-theoretic framework, we next assess how well the reconstructed posterior distributions recover the four spectral classes. For each held-out object, we compare its reconstructed posterior with the empirical latent distribution of each spectral class using KL divergence and assign the object to the class with the smallest divergence. Thus, the classification is based on distribution-level compatibility rather than on partitioning the latent space into fixed regions or assigning a class label from the posterior mean alone.

\begin{figure}
\centering
\includegraphics[width=1\columnwidth]{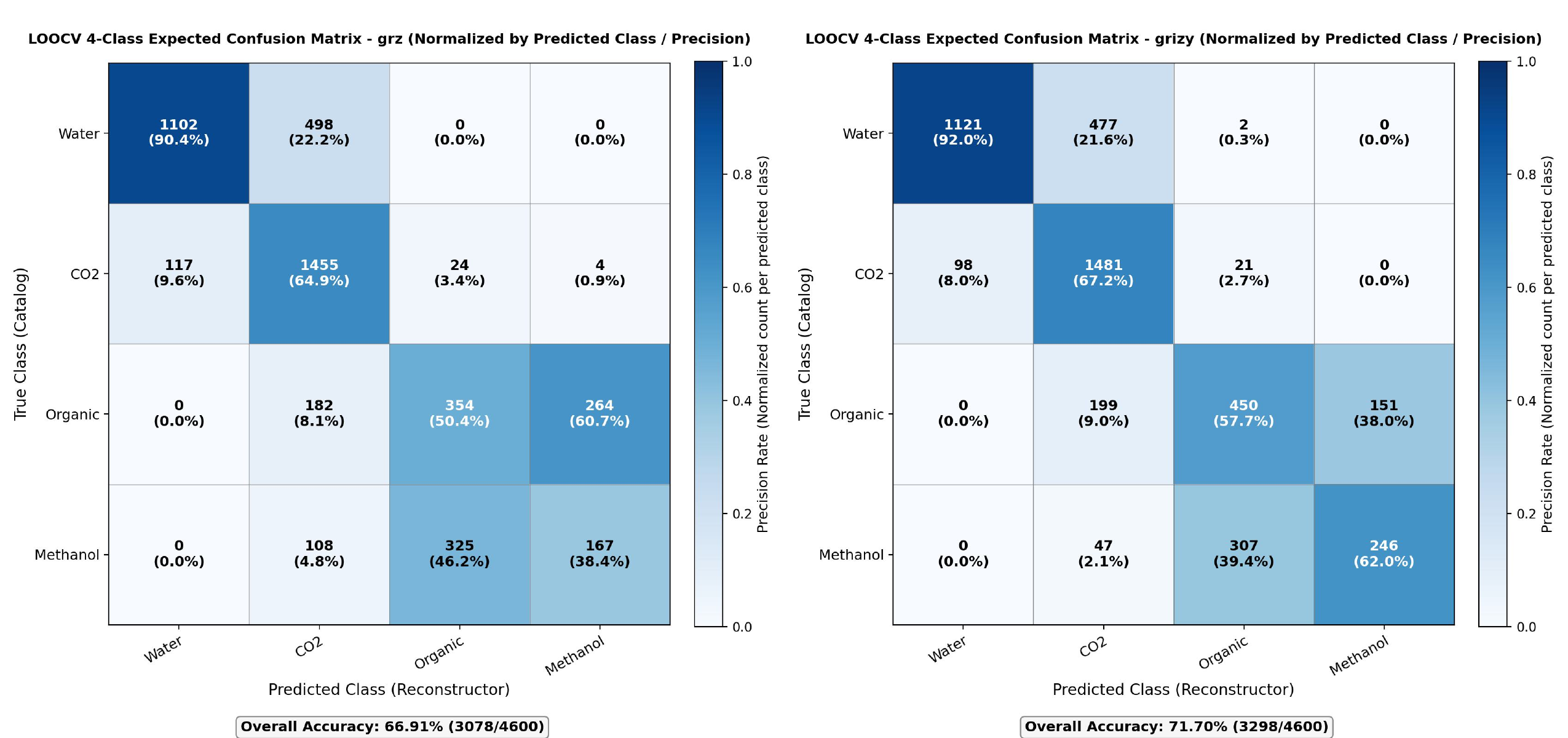}
\caption{
Leave-one-out cross-validation confusion matrices for the four-class TNO spectral taxonomy inferred from $grz$ (left) and $grizy$ (right) optical photometry. Each reconstructed posterior is assigned to the empirical spectral class with which it has the smallest KL divergence; no fixed classification boundaries are imposed in latent space. Rows indicate the true catalog classes, and columns indicate the reconstructed classes. Percentages are normalized within each predicted class and therefore represent classification precision; numbers show the summed counts across Monte Carlo realizations.}
\label{fig:cm}
\end{figure}

Figure~\ref{fig:cm} shows the resulting leave-one-out cross-validation confusion matrices for the $grz$ and $grizy$ filter sets. The modest increase in overall classification accuracy, from 66.9\% for $grz$ to 71.7\% for $grizy$, is consistent with the preceding information-metric analysis: three-band configurations such as $gri$ and $grz$ already capture most of the information required for broad spectral classification. If the organic-rich and methanol-rich subgroups are combined into a single broader organic-rich family, both the water-rich and organic-rich classes are recovered with high fidelity, whereas the CO$_2$-rich class is less cleanly separated. The improvement from extending the photometry to the $y$ band is mainly in resolving substructure within the organic-rich class, with the precision of methanol-rich predictions increasing from 38.4\% to 62.0\%.

These results show that optical colors are not simply redundant measurements of a single visible slope. Different filter combinations retain different amounts of information about the optical--NIR spectral manifold. This motivates the following analysis, in which optically defined color populations are interpreted as low-dimensional projections of the underlying spectral manifold.

\subsection{Colors to Spectral Manifold}
\label{sec:C2S}

We next ask how optically defined TNO populations project into the optical-NIR spectral manifold. Optical color populations can be defined in several ways, from hard color cuts \citep{Pike2017} and linear/nonlinear color-space transformations \citep{Fraser2023} to clustering \citep{Ferreira2025MNRAS}, and Gaussian-mixture models \citep{Bernardinelli2025}. The framework developed here is not tied to any particular optical classification scheme: any population described by optical color distributions can, in principle, be propagated into latent spectral space.

As a case study, we use the optical color groups identified by \citet{Bernardinelli2025}. Their Gaussian mixture model (GMM) separates the DES TNO color distribution into four subgroups, denoted NIRB+, NIRB-, NIRF+, and NIRF- (see Figure~\ref{fig:reverse_color_color} left panel). Because the populations are specified by a published Gaussian mixture model, they are fully reproducible from the model parameters alone, without requiring access to the original photometric sample. Furthermore, the groups were constructed entirely in optical color space and are independent of the spectral-manifold framework developed here. They therefore provide an independent benchmark for evaluating how empirical optical color populations are organized within the learned optical--NIR spectral manifold.

\begin{figure}
\centering
\includegraphics[width=.8\columnwidth]{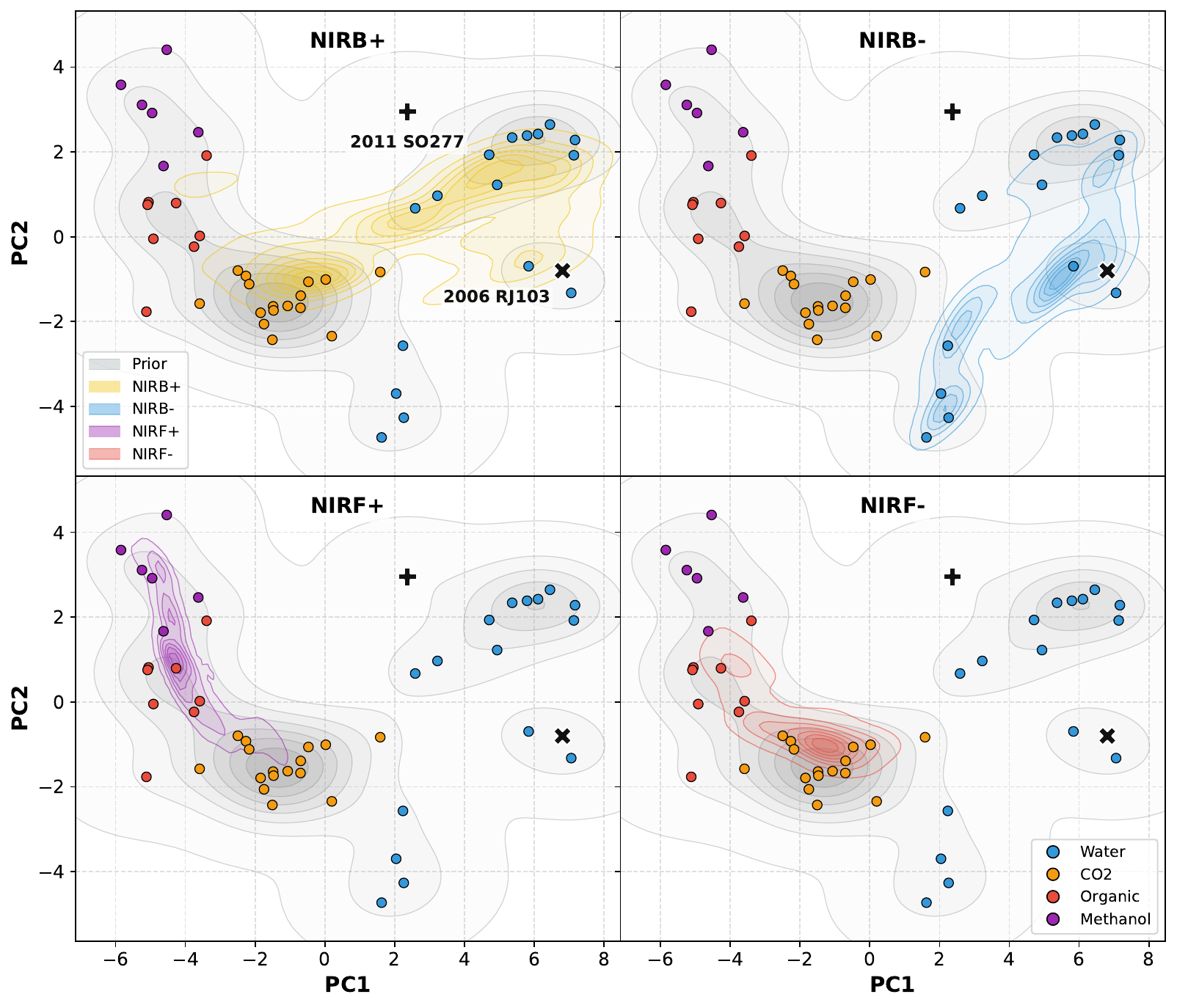}
\caption{
Projection of the Bernardinelli optical color groups into the optical-nearIR spectral latent manifold. Each panel shows the posterior latent distribution for one optically defined DES color group: NIRB+, NIRB-, NIRF+, and NIRF-. Gray contours show the KDE prior estimated from the training spectra, and colored points indicate the broad NIR spectral families in the JWST sample. The NIRF+ group maps to the red methanol-rich side of the manifold, NIRF- lies near the central CO$2$-rich and organic-rich region, NIRB- follows the water-side branch, and NIRB+ spans a broader transition region between the central and water-side branches. The reference objects 2011~SO$_{277}$ and 2006~RJ$_{103}$ are marked to illustrate the relation between optically selected groups and rare or transition-region spectral morphologies.
}
\label{fig:gmm_latent_projection}
\end{figure}

Figure~\ref{fig:gmm_latent_projection} shows the posterior latent distributions of the four optical groups in the PC1--PC2 plane. The gray contours show the empirical spectral prior, while the colored points indicate the training samples with the known NIR spectral classes. The four optical groups do not occupy random or identical regions of the latent manifold. Instead, they trace distinct branches and transition regions.
The clearest separation is between NIRB- and NIRF+. These two groups show (mostly) no overlap in latent space: NIRB- follows the water-side branch, whereas NIRF+ maps to the red organic-rich side of the manifold, including the methanol-rich subtype. This indicates that the blue-red optical separation captures a major compositional axis of the optical--NIR spectral manifold.

The other two groups, NIRB+ and NIRF-, occupy a more intermediate region and overlap most strongly near the CO$_2$-rich branch. This is consistent with their proximity in optical color-color space, but the latent projection gives this proximity a spectral interpretation: both groups sample the central part of the optical-NIR manifold, where CO$_2$-rich, organic-rich, and water-side morphologies approach one another. NIRB+ extends farther toward the water-side branch, whereas NIRF- remains closer to the central red / CO$_2$-rich region.

%The locations of two spectrally distinct Neptune Trojans, 2011~SO${277}$ and 2006~RJ${103}$, provide useful reference points. 2011~SO${277}$ lies near the intermediate region sampled by NIRB+, while 2006~RJ${103}$ lies in a distinct water-side, low-density region covered by NIRB-. This comparison shows that spectrally distinct objects can occupy low-probability regions of the latent distributions.

\begin{figure}
\centering
\includegraphics[width=.7\columnwidth]{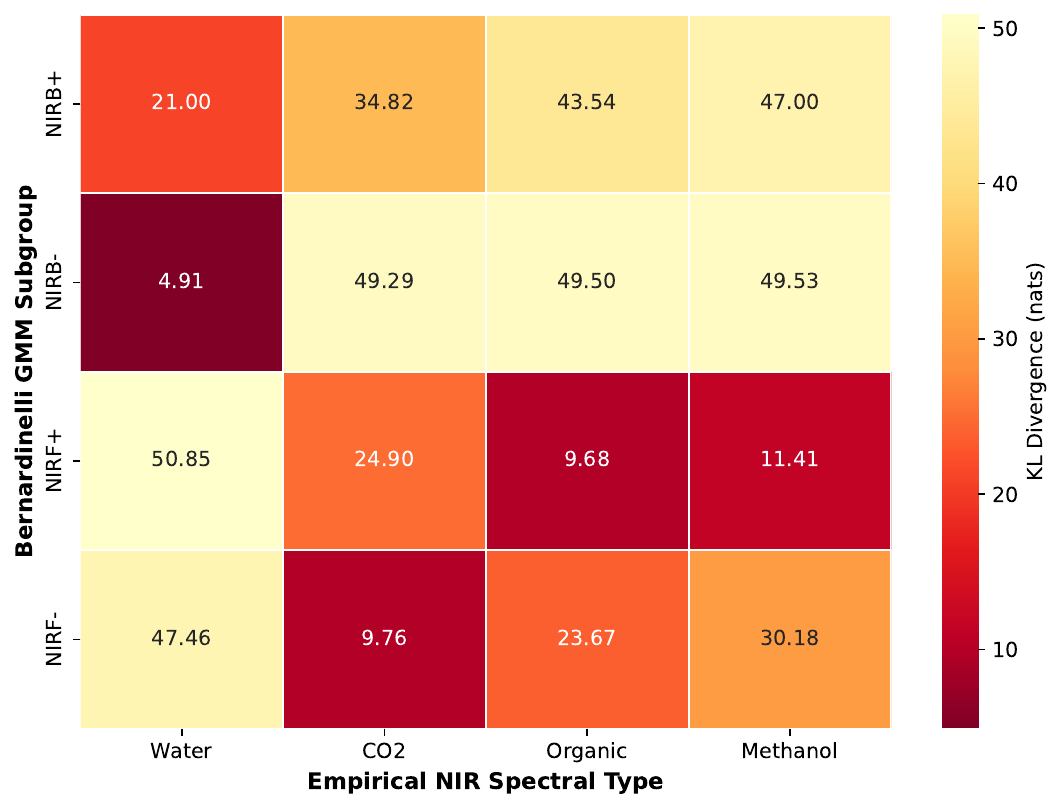}
\caption{KL divergence between the \citet{Bernardinelli2025} optical GMM subgroups and the empirical NIR compositional branches. Lower values indicate a closer match between the optical GMM subgroup and the corresponding NIR spectral family. Rather than comparing objects individually, the KL divergence quantifies the similarity between the posterior distributions of each population in the learned spectral manifold.}
\label{fig:gmm_kl_matrix}
\end{figure}

To quantify these visual impressions, Figure~\ref{fig:gmm_kl_matrix} compares each optically defined group with the empirical near-infrared spectral families using latent-space KL divergence. Lower values indicate that the optical group distribution is more compatible with the corresponding NIR spectral family. The matrix confirms that NIRB- is strongly associated with the water-rich family, whereas NIRF- is closest to the CO$_2$-rich family. NIRF+ is closest to the organic-rich and methanol-rich side of the manifold, consistent with its red optical colors and latent-space location. NIRB+ is also closest to the water-rich family, but with a substantially larger KL divergence than NIRB-, reflecting its broader and more transitional latent distribution. These results show that the Bernardinelli optical color groups are not just statistical clusters in color-color space. When interpreted through the optical-nearIR spectral manifold, they appear as low-dimensional projections of chemically structured spectral classes and transitions. 
%We emphasize that these KL values should not be only interpreted as classification probabilities. They are information distances between distributions, and are therefore most useful for comparing relative compatibility between optically defined populations and empirical NIR spectral families.

\begin{figure}
\centering
\includegraphics[width=.8\columnwidth]{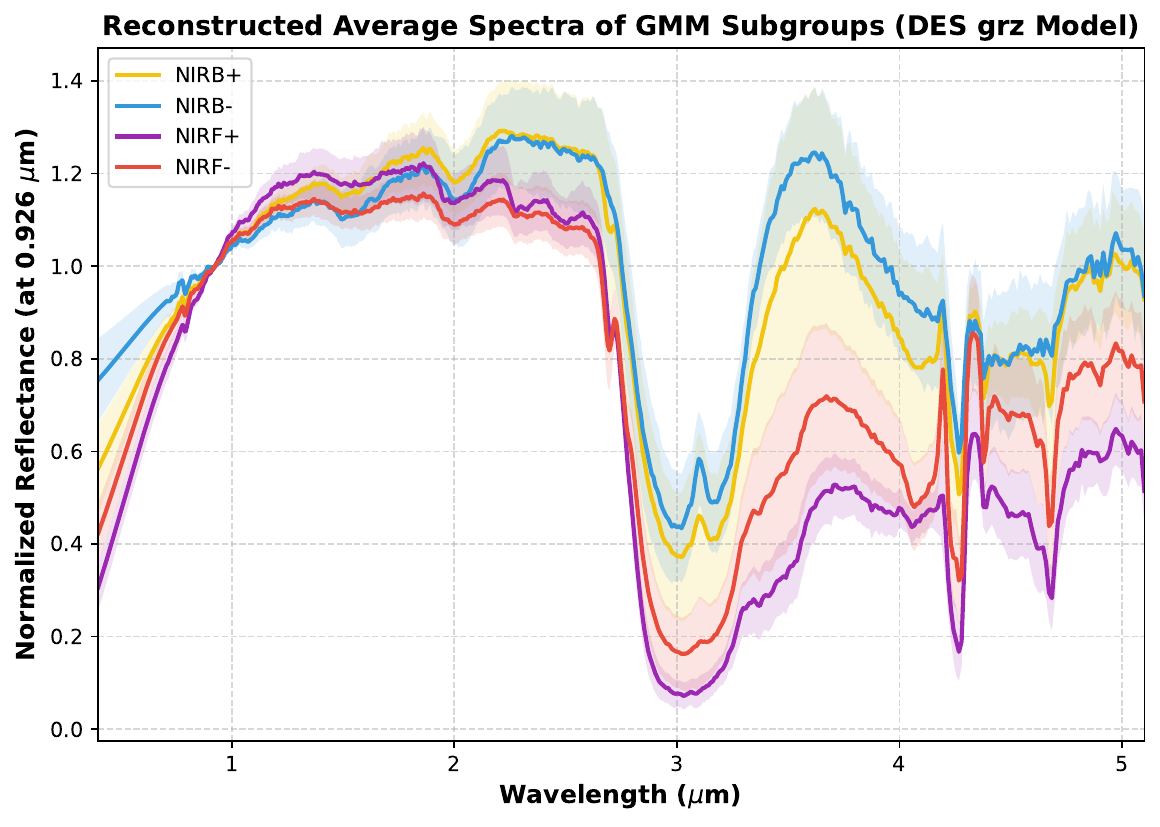}
\caption{
Posterior predictive spectra for the four Bernardinelli optical color groups, reconstructed using the DES grz model. Each curve shows the mean decoded spectrum of posterior samples for a given optical group, normalized at $0.926~\mu{\rm m}$. Shaded regions show posterior predictive intervals and represent the range of spectra compatible with each optically defined population, not measurement uncertainties on individual spectra. NIRF+ maps to the reddest and most strongly absorbed spectral morphology, NIRF- occupies a less extreme red / central branch, and the two NIRB groups are associated with the water-side part of the manifold. NIRB+ shows broader uncertainty and a more intermediate morphology than NIRB-, consistent with its extended latent-space distribution.
}
\label{fig:gmm_average_spectra}
\end{figure}

Figure~\ref{fig:gmm_average_spectra} shows the corresponding posterior predictive spectra for the four optical color groups. This wavelength-space view confirms the separation seen in latent space. NIRF+ produces the reddest and most strongly absorbed spectrum, consistent with its location on the methanol-rich side of the manifold. NIRF- is also red in the optical, but its NIR morphology is less extreme and more closely resembles the central CO$_2$-rich / organic-rich classes. The two NIRB groups show comparatively bluer optical continua and stronger similarity to the water-types, but they are not identical. NIRB- produces a relatively coherent water-side spectrum, whereas NIRB+ shows broader uncertainty and a more intermediate morphology, consistent with its extended latent-space distribution.
The shaded regions in Figure~\ref{fig:gmm_average_spectra} represent posterior predictive intervals induced by the latent distributions of each optical group. They should therefore be interpreted as the range of spectra compatible with the optically defined population, not as measurement uncertainties on individual spectra. The larger uncertainty for NIRB+ reflects its broader latent-space support and reinforces the interpretation that this group spans a transition region rather than a single narrow spectral phenotype.

\subsection{Spectral Manifold to Colors}
\label{sec:S2C}

Having used optical colors to infer posterior distributions over the spectral manifold, we now examine the inverse direction by generating the optical color distribution from the learned spectral manifold. For each of the four spectroscopically defined compositional groups, H$_2$O-rich, CO$_2$-rich, organic-rich, and methanol-rich, we sample their empirical densities in PCA latent space, decode each sample into a synthetic optical---NIR reflectance spectrum, and compute the corresponding $g-r$ and $r-z$ colors. Unlike previous analyses that characterize the observed optical color distribution directly, the optical populations shown here emerge naturally from the learned spectral manifold through the forward observation model. The resulting color distributions therefore represent a generative interpretation of the observed optical color space rather than a statistical smoothing of the measured colors.

\begin{figure}
\centering
\includegraphics[width=1\columnwidth]{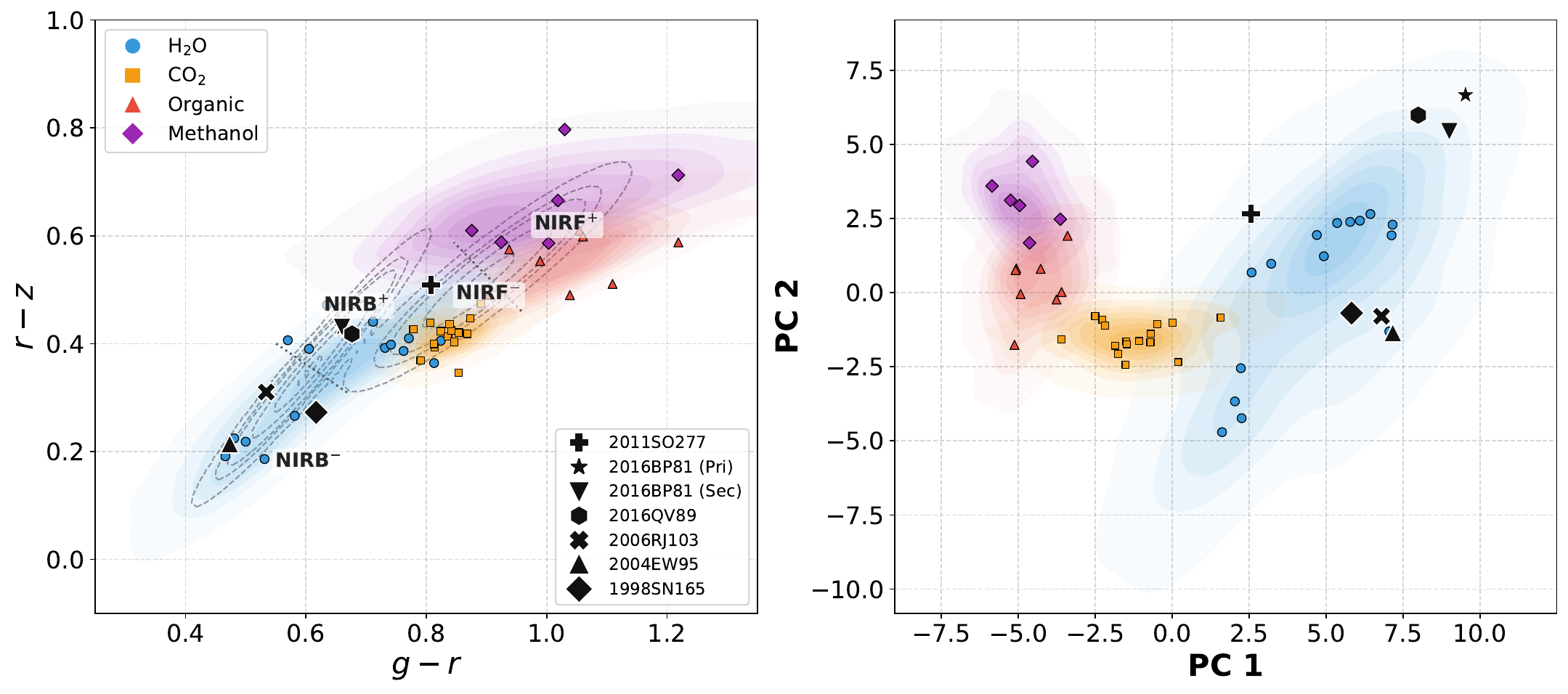}
\caption{Generating Optical Color Distributions from the Spectral Manifold. Left: Synthetic DES $g-r$ and $r-z$ colors calculated from the spectra decoded from these latent samples. Colored density contours represent the generated distributions of the four spectral groups, and gray dashed contours indicate the empirical NIRB and NIRF optical populations from \citet{Bernardinelli2025}. Several spectrally unusual objects are overplotted for comparison. Right: Samples drawn from the empirical distributions of the four spectroscopically defined compositional groups (H$_2$O-rich, CO$_2$-rich, organic-rich, and methanol-rich) in the first two principal components of the spectral latent space. Colored symbols denote the training objects with measured optical colors and NIR spectra.
}
\label{fig:reverse_color_color}
\end{figure}

Figure~\ref{fig:reverse_color_color} illustrates the generative mapping from the learned spectral manifold to optical color space. The right panel shows samples drawn from the first two principal components of the spectral manifold, while the left panel displays the corresponding synthetic optical colors obtained by decoding these latent samples into spectra and computing their broadband colors. Samples from the H$_2$O-rich class populate the bluest region of optical color space, the CO$_2$-rich class occupies an intermediate and relatively compact locus, while the organic-rich and methanol-rich classes extend toward progressively redder colors. The generated color distributions are distinct but partially overlapping, demonstrating that optical colors preserve the large-scale topology of the spectral manifold while compressing its higher-dimensional structure.

The gray dashed contours show the empirical NIRB and NIRF optical populations from \citet{Bernardinelli2025}. Their close agreement with the generated optical color distributions supports the interpretation that the observed optical populations are low-dimensional manifestations of the continuous optical--NIR spectral manifold, rather than independent empirical classes. The overlap among the generated distributions also explains why broadband optical colors, despite retaining substantial compositional information, cannot uniquely recover detailed near-infrared spectral types. Additional near-infrared observations are therefore required to resolve objects that occupy transition regions of the manifold.

\section{Discussion} \label{sec:dis}

\subsection{The Information Content of Broadband Photometry}

Broadband optical colors have long been used as proxies for TNO surface composition. Until recently, however, it has remained unclear how much spectral information survives the compression from a continuous optical--NIR spectrum into only a few broadband measurements. The availability of homogeneous JWST spectroscopy now makes it possible to address this question quantitatively by constructing an empirical optical--NIR spectral manifold.

Our results suggest that broadband optical photometry should be viewed as a lossy but information-rich encoding of this manifold. Although high-frequency spectral details are inevitably discarded, the dominant compositional geometry is largely preserved, allowing major spectral classes to remain distinguishable from optical colors alone. In this framework, spectroscopy and broadband photometry are not competing approaches but complementary representations of the same compositional manifold at different information resolutions: JWST spectroscopy defines the manifold, while wide-field optical surveys provide compressed observations that localize objects within it. This perspective naturally motivates the generative interpretation developed in the following section.

\subsection{The Optical Color Distribution as a Projection of the Spectral Manifold}

The central hypothesis of this work is that broadband optical colors represent compressed optical--NIR spectral information. If this hypothesis is correct, then a spectral manifold learned from JWST spectroscopy should naturally reproduce the observed topology of optical color space. Figure~\ref{fig:reverse_color_color} provides this validation through a generative mapping from the spectral manifold to optical color space. By sampling the learned latent distribution, decoding the corresponding spectra, and computing their broadband colors, the model reproduces the principal structures observed in color--color space without fitting the optical color distribution directly. The generated distributions recover the broad separation between the water-rich, CO$_2$-rich, and organic-rich populations, as well as their partial overlap.

This perspective provides a unified physical interpretation of previous optical color studies. The color--color structures reported by \citet{Pike2017}, \citet{Schwamb2019}, \citet{Fraser2023}, \citet{Bernardinelli2025}, \citet{Ferreira2025MNRAS}, and related works need not be viewed as isolated empirical classifications. Instead, they can be understood as different observational manifestations of the same underlying optical--NIR spectral manifold, sampled with different surveys, filter systems, and statistical methodologies.
Earlier studies successfully characterized the topology of optical color space from photometric data alone, whereas the present work provides a physical interpretation of that topology by linking it to the underlying spectral manifold. Broadband optical colors are therefore more than empirical observables to be clustered; they are compressed representations of the optical--NIR compositional manifold that preserve much of its large-scale geometric structure.

\subsection{Implications for Survey-Era TNO Surface Science}

While the reconstruction of individual spectra provides a useful validation of the framework, its primary scientific value lies in enabling population-scale compositional studies from broadband optical surveys. By statistically linking optical colors to the optical--NIR spectral manifold, wide-field surveys such as Rubin Observatory's LSST will enable compositional maps of the outer Solar System, allowing the distribution of spectral-manifold posteriors to be investigated as a function of dynamical class, heliocentric distance, inclination, resonance, and other orbital parameters.

Although broadband optical colors cannot uniquely determine every spectral type, they provide an efficient means of localizing objects within the spectral manifold and identifying candidates for spectroscopic follow-up. Figure~\ref{fig:reverse_color_color}, for example, suggests that candidates with weak-water spectra similar to 2006~RJ$_{103}$ are expected to preferentially occupy the bluest NIRB- population, whereas ``blue-binary''-like spectra \citep[2016~QV$_{89}$, 2016~BP$_{81}$,][]{Wong2025PSJ} may preferentially appear within the NIRB+ region. In contrast, analogs of the unusual spectrum represented by 2011~SO$_{277}$ are expected to lie near the overlap between multiple optical populations, making these transition regions particularly attractive targets for follow-up spectroscopy.

\begin{figure}
\centering
\includegraphics[width=1\columnwidth]{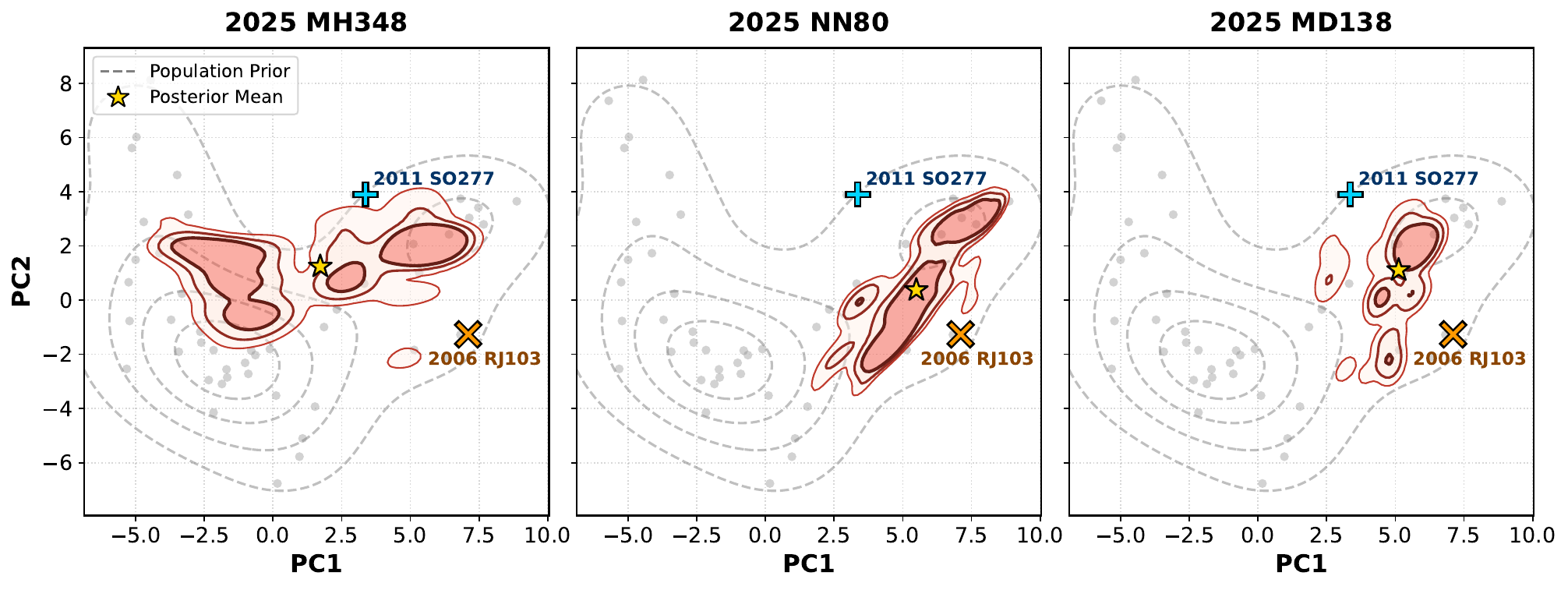}
\caption{Latent PCA space (PC1 vs PC2) posterior probability distributions for the three Neptune Trojans, 2025~MH$_{348}$, 2025~NN$_{80}$, and 2025~MD$_{138}$, reconstructed from optcial colors. The gray points and dashed background contours represent the empirical reference population manifold derived from high-resolution TNO spectra (TNO Population Prior). The red shaded contours correspond to the $1\sigma (68.3\%), 1.5\sigma (86.6\%)$, and $2\sigma (95.4\%)$ posterior credibility regions derived via 2D Gaussian Kernel Density Estimation (KDE) across 10,000 Monte Carlo realizations propagating observational photometric uncertainties. Gold star markers indicate the posterior mean latent coordinates for each object.
}
\label{fig:sv_nt_latent}
\end{figure}

This predictive capability is demonstrated using Rubin/LSST multiband observations of three L5 Neptune Trojans discovered in the Rubin Science Validation (SV) Survey \citep{MPS2026, claver2025, guy2026}. Figure~\ref{fig:sv_nt_latent} shows that the posterior distributions inferred from the broadband optical colors reported by \citet{Schwamb2026} occupy different regions of the spectral manifold. Two objects (2025~NN$_{80}$ and 2025~MD$_{138}$) are tightly localized within the canonical water-type class, whereas 2025~MH$_{348}$ exhibits a broader posterior extending toward the CO$_2$-rich region. Given that no confirmed CO$2$-rich Neptune Trojan is currently known, and that 2011~SO$_{277}$ remains the closest analog to such a transition spectrum \citep{Markwardt2025}, the posterior distribution of 2025~MH$_{348}$ suggests that it may occupy a similar region of the spectral manifold, making it an attractive target for NIR photometry/spectroscopic follow-up.

Finally, the spectral manifold should not be regarded as static. As additional high-quality JWST spectra become available, particularly for currently underrepresented or unusual surface types, the manifold will become increasingly complete and better constrained. Rather than defining a fixed taxonomy, the spectral manifold is expected to evolve into an increasingly complete representation of the continuous compositional diversity of the trans-Neptunian population.

\section{Summary}\label{sec:sum}

In this work, we presented a probabilistic framework that links broadband optical photometry to an optical-NIR spectral manifold constructed from JWST spectroscopy. By combining Bayesian inference with a low-dimensional manifold representation, we show that broadband optical colors preserve substantial compositional information despite representing a highly compressed view of the underlying spectra. Information-theoretic metrics further provide a quantitative measure of the uncertainty and information retained by different photometric filter sets.

The learned spectral manifold enables inference in both directions. Optical colors constrain posterior distributions on the manifold, while generative sampling from the manifold naturally reproduces the observed topology of optical color space. This demonstrates that previously identified optical color populations are not merely empirical statistical groupings, but observational manifestations of a continuous optical--NIR compositional manifold.

The framework also provides a practical foundation for survey-era Solar System science. As Rubin Observatory's LSST dramatically expands the number of trans-Neptunian objects with precise broadband photometry, the learned spectral manifold enables population-scale compositional mapping, efficient identification of rare or transitional surface types for spectroscopic follow-up, and quantitative evaluation of the information content of future photometric surveys. As additional JWST spectra become available, the learned manifold itself will continue to evolve into an increasingly complete representation of outer Solar System compositional diversity.

\appendix
\restartappendixnumbering

\section{The Linear Continuum Model in the Spectral-Manifold Framework}
\label{app:linear_continuum}
\begin{figure}
\centering
\includegraphics[width=1\columnwidth]{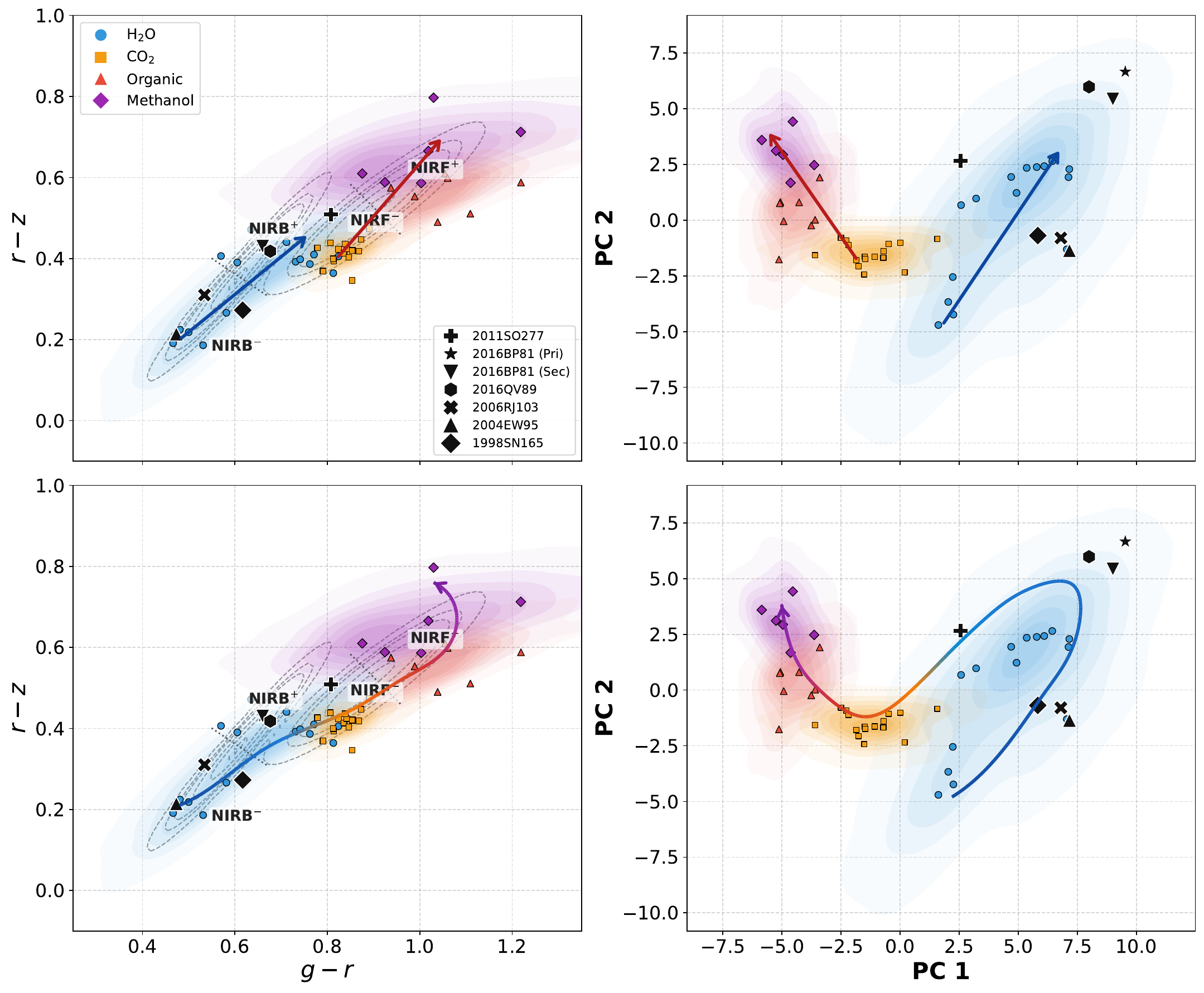}
\caption{Schematic illustration of the relationship between the linear continuum model of \citet{Fraser2026} and the spectral-manifold framework developed in this work. The upper panels illustrate the approximately one-dimensional representation of the two major spectral groups, water-rich and CO$_2$--organic types, used to describe the dominant optical color--spectral relations in the linear continuum model. The lower panels illustrate the corresponding continuous, nonlinear structure represented by the spectral manifold in this work. The figure provides a conceptual interpretation of the linear continuum relations as low-dimensional representations of the broader spectral structure, rather than a formal mathematical derivation of the two models.
}
\label{fig:appendix}
\end{figure}

Independent of this study, a linear continuum relation between optical color slope and TNO spectra was proposed by \citet{Fraser2026}. The Fraser model and the present work share the same basic premise: TNO spectra exhibit low-dimensional structure. Here, we provide a possible interpretation of the linear model within our spectral-manifold framework.
Figure~\ref{fig:appendix} illustrates this correspondence schematically. The upper two panels illustrate how the optical color--spectral relation can be represented by a linear model. In this representation, the spectral manifold is effectively separated into two major groups corresponding approximately to water-rich and CO$_2$--organic spectral types, as identified by \citet{Fraser2026}. The lower two panels illustrate how the same spectral structure is represented as a continuous, nonlinear manifold in our framework.

\section*{Software and Data Availability}

The complete optical-color-to-spectral inference framework is fully open source and publicly available on GitHub at \url{https://github.com/sevenlin123/color_to_spec}. The repository provides Python source code and an AI-agent skill implementing the complete inference workflow, including the trained PCA spectral manifold and KDE prior, Bayesian posterior inference, information-theoretic metrics, probabilistic compositional inference, spectrum generation, and generative optical color synthesis. Following the reproducibility framework of \citet{Lin2026Reproducibility}, the agent skill serves as an executable implementation of a reproduction prompt set, providing a structured and machine-readable representation of the workflow for reproducible execution by AI agents.

\begin{acknowledgments}

%The authors are grateful to the anonymous referee for the thorough review and highly constructive feedback. 
The authors thank Noem\'{\i} Pinilla-Alonso and Rosario Brunetto for kindly providing us the reduced TNO spectra. H.W.L. thanks Ying-Tung Chen for comments on an earlier draft and Zong-Fu Sie for helpful discussions and for identifying bugs in the code. As H.W.L's final paper completed while at the University of Michigan, this work is offered with gratitude to both the institution and to Claude Shannon -- a UM alumnus and the father of information theory -- whose ideas underpin the entropy and divergence measures central to this analysis.

All of the data presented in this paper were obtained from the Mikulski Archive for Space Telescopes (MAST) at the Space Telescope Science Institute. 
STScI is operated by the Association of Universities for Research in Astronomy, Inc., under NASA contract NAS5-26555. Support to MAST for these data is provided by the NASA Office of Space Science via grant NAG5-7584 and by other grants and contracts. 
Support for program \#2550 was provided by NASA through a grant from the Space Telescope Science Institute, which is operated by the Association of Universities for Research in Astronomy, Inc., under NASA contract NAS 5-03127. The authors acknowledge the support of the National Science Foundation under Grant No. AST-2406527. 
L.M. was also supported by the Marsden Fund Council from Government funding, managed by Royal Society Te Apārangi. R.M. acknowledges support from NASA under Agreement No. 80NSSC21K0593 for the program “Alien Earths”.

Statement on the use of generative AI tools: The authors disclose the following use of generative-AI tools in the preparation of this manuscript: Code agent (Google Antigravity): 1. Python scripts for producing the figures. 2. boiler-plate loops and batch-execution scripts for non-critical parameter/sample-sweep experiments. 
English language editing (OpenAI Prism): Drafting and polishing of the English text.
These AI tools were used solely as assistance and are not listed as co-authors. All scientific judgments, experimental design, and interpretation of results remain the responsibility of the authors.
\end{acknowledgments}

\vspace{5mm}
\facilities{JWST (NIRSpec)}


\software{AutoGluon \citep{agtabular}, astropy \citep{2013A&A...558A..33A,2018AJ....156..123A,astropy2022}, scipy \citep{2020SciPy-NMeth}, scikit-learn \citep{scikit-learn}, jwst \citep{2023zndo...8247246B}}
%% For this sample we use BibTeX plus aasjournalv7.bst to generate the

\bibliography{sample701}{}
\bibliographystyle{aasjournalv7}



\end{document}